\documentclass[submission,copyright,creativecommons]{eptcs}
\usepackage{breakurl}             % Not needed if you use pdflatex only. 

\providecommand{\event}{FM \& MDD 2017} % Name of the event you are submitting to
\usepackage{amsmath, amssymb}
\usepackage{graphicx}
\usepackage{url}
\usepackage{capt-of}
\usepackage{listings}
\usepackage{color}
\usepackage{hyperref}
\hypersetup{pdftex,colorlinks=true,allcolors=blue}
\usepackage{hypcap}
\usepackage{multirow}
\usepackage[colorinlistoftodos]{todonotes}

\lstdefinelanguage{REST} {morekeywords={precondition,postcondition,resource,method,PATH,POST,GET,PUT,DELETE},}

\title{Securing Open Source Clouds Using Models}

\author{Irum Rauf
\institute{\AA bo Akademi University, Turku, Finland}
\email{irum.rauf@abo.fi}
\and
Elena Troubitsyna
\institute{\AA bo Akademi University, Turku, Finland}
\email{\quad elena.troubitsyna@abo.fi}
}
\def\titlerunning{Securing Open Source Clouds Using Models}
\def\authorrunning{Irum Rauf \& Elena Troubitsyna }

\begin{document}
 \maketitle

 \begin{abstract}  
 The widespread adoption of cloud computing has resulted in proliferation of open source cloud computing frameworks that give more control to enterprises over their data and networks. Though, the benefits of the open source software are widely recognized, there is a growing concern over their security assurance. Often open source software is a subject of frequent updates. The updates might introduce or remove a variety of features and hence violate security properties of the previous releases. Obviously, a manual inspection of security would be prohibitively slow and inefficient. In this work, we propose an automated approach that can help developers to assure security of open source cloud framework even in the presence of frequent releases. Our methodology consists of creating a (stateful) wrapper that emulates the usage scenarios with explicit representation of security and functional requirements as contracts. We use a model-driven approach to model REST APIs of KeyStone, an identity service in OpenStack. Openstack is an open source cloud computing framework providing IaaS. Our models define structural and behavioral properties of Keystone together with its security requirements. We detail the implementation of these models in Django Web Framework and also show how to use the behavioral interfaces to implement a service monitor for the cloud services. This mechanism facilitates verification and validation of functional behavior and security requirement in an automated manner. 

 \end{abstract}

\section{Introduction}
Open source cloud frameworks allow their customers to build their own private Infrastructure as a Service (IaaS). IaaS provides Virtual Machines (VMs) under the pay-per-use business model \cite{mell2011nist}. The source code of Open Source (OS) clouds is distributed publicly. Moreover, often software is developed in a collaborative manner that makes it  a subject of frequent updates. These updates might introduce or remove a variety of features and hence, violate the security properties of the previous releases.

Assuring the security of opensource clouds is an important concern for cloud providers. Often open source clouds use REST architectural style to offer their APIs. REST offers a different architectural style to invoke remote services in contrast to contemporary SOAP-based services. Its different architecural style motivates the need to develop novel design and security assurance methodologies to handle its stateless protocol for developing stateful services. Stateful services can have different states that a service must go through during its lifecycle. It requires a certain sequence of method invocations that must be followed in order to fulfill the functionality a service promises to deliver to its users. In this work, we propose a methodology that consists of creating a (stateful) wrapper that emulates the usage scenarios and contains an explicit representation of security and functional requirements as contracts.

We adopt a model-driven approach -- \underline{Se}curity and \underline{Re}st compliant \underline{UM}L Models (SecReUM) -- that builds on the theory presented in \cite{porres2011modeling} to create a security-validating wrapper. We define the structural interface of a REST API using UML class diagram. The usage scenarios -- the dynamic behaviors --   are represented as state diagrams. These models lead to RESTful interfaces, describe the behavior of operations in terms of preconditions and postconditions and also facilitate the specification of the authentication mechanism. In this work, we demonstrate how to generate contracts defining the security properties as pre- and post-conditions using these models and implement them as a wrapper for the cloud implementation.
% We use contracts with models to provide \underline{Se}curity and \underline{Re}st compliant \underline{UM}L Models (SecReUM), 
% By using model-based test generation approach, we can generate test cases from SecReUM that can validate the behavior of software. SecReUM can also be used to provide an online/offline monitoring mechanism for KeyStone.

The approach is implemented as a wrapper in Django Web Framework \cite{holovaty2009definitive} for the KeyStone component of OpenStack.
OpenStack is an open-source software platform for cloud computing that offers REST interfaces to provide IaaS (Infrastructure as a Service)\cite{sefraoui2012OpenStack}.  Keystone offers identity service in OpenStack for authentication and authorization. 

%The security assurance of open source clouds is an important concern Open source clouds with REST APIs requires novel design and validation approaches since REST APIs require usage of design methodologies and security mechanisms that can handle stateless protocol for stateful applications.  becomes  is combined with REST architectural style. The adoption of REST architecture brings additional benefits of scalability and extensibility and encourages providers to offer their services to a wider audience as well as add more features.    
The paper is organized as: Section 2 briefly explains Keystone and its interface. Section 3 presents an overview of our overall approach. Section 4 explains our modeling approach for REST APIs with stateful behavioral and section 5 shows the generation of contracts with security concerns. The implementation of the approach is presented in section 6. Section 7 and section 8 show the applications of our work and the related work, respectively. The paper is concluded in section 8.

\section{Keystone Open Stack}

Keystone  is the centralized identity service of OpenStack that offers authentication and authorization \cite{sefraoui2012OpenStack}. KeyStone authenticates a user by generating a token. A token can either be scoped or unscoped depending on the client's request and the configured policy of KeyStone. An unscoped token authenticates a user without authorising for any project.  In contrast, a scoped token provides the authorization information of the user for a particular project or domain.  

 KeyStone offers REST API in compliance with OpenStack policy \cite{ksapi}. REST services expose their functionality as resources and each resource has a unique URI that provides \emph{addressability}. CRUD (create, retrieve, update and delete) operations can be performed on resources using standard HTTP methods. This means that only HTTP request methods (GET, PUT, POST, DELETE) can be invoked on KeyStone resources. In order to offer scalability, the \emph{statelessness} feature of REST is ensured by treating every request independently without requiring any session or cookie information from user requests.  Each resource, when invoked via its URI and a standard HTTP method, replies with a status code and a resource representation, which contains the data about the resource attributes and links to other resources. The HTTP response code is a numeric code that tells the clients whether the request went successfully. HTTP has a list of status codes that reveal how the request went \cite{berners1996hypertext}, for example,  200 means the request was successful, 404 means the resource was not found and 403 implies that it is forbidden to make this request on this resource. The client machine interpret these response codes to know how their request went.

\section{Overall Approach}
 
%Open-source cloud environments are updated frequently by different users. It becomes challenging for the providers to validate that the software continues to comply with its functional and security requirements after the updates. Typically, the in-house software/ security team manually look for changes and run different types of analysis techniques, ranging from manual code-inspections to testing, to identify errors. Our work enables automated security monitoring and validation by establishing a model-driven security assurance framework for open-source clouds. The framework allows the providers of open-source cloud to periodically validate their cloud services for the functional and security requirements it promises to deliver. 

\begin{figure}[h]
\centering
\includegraphics[width=11cm]{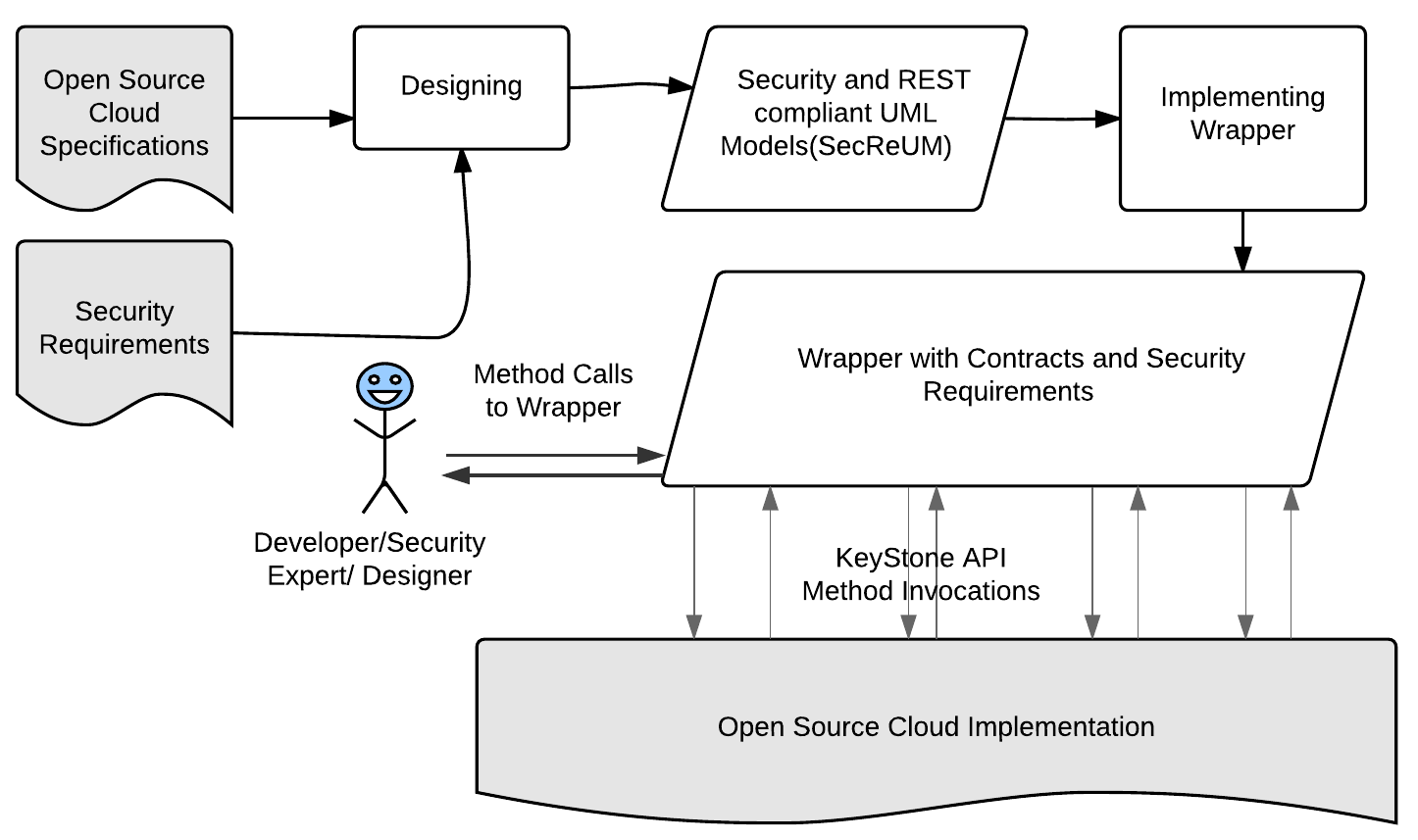}
\caption{Model-Driven Framework for Security Assurance}
\label{fig:mdf}
\end{figure}

 In this section we give an overview of our overall approach to create security-validating wrapper. The approach is presented in Figure \ref{fig:mdf}. It consists of two main steps 1) Designing SecReUM and 2) Implementing wrapper with behavioral contracts. 

 The specifications and implementation of the open source cloud frameworks, that are publicly available, are taken as input. The security requirements for the system are provided by security experts and also taken as an input. These three entities are shown as grey boxes in Figure \ref{fig:mdf} to indicate their availability beforehand. 

In the first step, Security and REST compliant UML Models (SecReUM)  are designed using our approach detailed in Section \ref{sec:modeling1}. 

In the second step, we build upon the design by contract strategy and generate contracts that define the security properties from SecReUM that are implemented as code skeletons in the wrapper program.
 The code-skeletons can be generated using our tool presented in \cite{rauf2011beyond}. The tool generates code skeleton for design models in Django that is a high level Python web framework \cite{djbook}.   

 A wrapper program is capable of invoking another program, perhaps with a larger body of code,  by providing an interface to call. Our code skeletons, generated for the security-validating wrapper, has behavioral information, i.e., contracts for each method and the developer just has to write the implementation of the operations, i.e., invocation to the cloud implementation. Contracts use pre- and post-conditions for the methods to define correctness conditions of the program. They are capable of detecting a change in the state of the program and identifying when a certain piece of code violates the pre-defined conditions. Moreover, they can be used for fault localization. Our approach builds upon these features offered by contracts and invoke the cloud implementation through our wrapper program to validate if the cloud implementation conforms to its security specifications.  

 Our approach can be used by an in-house developer, a designer of the cloud of an organization or a security expert to validate if the implementation of the cloud is as specified. As shown in Figure \ref{fig:mdf}, user invokes the wrapper with the right method call. The wrapper program checks the user request and invokes the cloud implementation only if the pre-conditions for that method are satisfied. The wrapper, thus, constraint the user to invoke the service under the right conditions. This information can be used by security experts to know whether a method can be invoked on the cloud implementation if different resources of the service are not in the specified or required state. Similarly, the response from the cloud implementation is received by the wrapper and checked to see if it satisfies the post-condition for that method. The corresponding response to the user is only given if the post-condition for that method is satisfied. The post-condition, thus, constraint the implementation to provide the correct functionality and fulfill the security concerns expected from it. This gives security experts the information whether a feature is removed or updated in the cloud implementation that was not intended during the update resulting in functional incorrectness or compromise on the expected security features.

%The third step of our framework is \emph{Testing} in which test cases are generated using different model-based test generation approaches from  SecReUM. These test cases are run against the wrapper program to validate the implementation of open source software. Thus, by periodically running the same test suites (in case the specifications are unchanged) or updated one (in case the specifications are changed), the implementation of open source software can be validated and errors can be identified using pass/fail results of the test cases. 

% The rest of the paper discusses on details of the proposed approach. Section \ref{sec:modeling1} focuses on designing steps, section \ref{sec:contract} and \ref{sec:imp} explain contract generation and implementation correspondingly. 
%  %In this paper, we focus in detail on our designing, contract generation and implementation approach, presented in section \ref{sec:modeling1} section \ref{sec:contract} and section \ref{sec:imp}, respectively. 
%  The model-based test generation from SecReUM is out of the scope of this paper and hence not addressed. However, for validation, we can not only benefit from our previous work for validating behavioral REST interfaces \cite{rauf2014scenario} but can also take advantage of the large body of work done in generating test cases from behavioral contracts using UML as a familiar notation.

\section{Modeling approach for SecReUM}
 
\label{sec:modeling1}
REST APIs use stateless protocol but they can be used to create applications with complex stateful behavior. Stateful behavior requires that methods are invoked in a particular sequence to fulfill a specific goal. For example, in order to delete a user in KeyStone, the user must first authenticate herself in \emph{admin} role and also get a valid scoped token. The stateless feature of REST implies that every method is treated independently without requiring information about the methods invoked earlier by not keeping any hidden state or session information. All the methods in REST API are self-contained, i.e., all the information required to invoke the method is contained in the invoked method. By adopting this architecture style, scalable services can be offered. 
However, in addition to preserving sequence of method calls, stateful behavior also offers information about the conditions under which these methods should be invoked in order to fulfill service goals. This information can be used in a variety of ways in order to determine if the service continues to offer the functional and non-functional properties it promises to deliver. 
In this section, we present our approach to model stateful services using REST architecture style.

%for REST APIs offe
%The benefit of giving a stateful view to this behavior of KeyStone facilitates the understanding of KeyStone behavior and helps in validating the functional and non-functional behavior of KeyStone by defining conditions under which different methods can be invoked.

In section \ref{sec:rm} and section \ref{sec:bm}, we  present resource model as UML class diagram and behavioral model with UML state machine, respectively, with additional constraints to represent REST features. Figure \ref{fig:ks_rm} and Figure \ref{fig:ks_bm} gives an excerpt of REST interface of KeyStone as an example. 

%We model the behavioral interface of KeyStone from the viewpoint of our wrapper program that will invoke the KeyStone and can constrain the user to invoke the service under right conditions and service provider to fulfill the functionality expected from it. \\

\subsection{Resource Model:} 

\label{sec:rm} 
 A UML class diagram represents the classes of a software and the associations between them. An association defines a relationship between two classes by which one class knows about the other class~\cite{uml20112}. We are using UML class diagram with additional design constraints to represent resources, their properties and relation with each other.  The concept of a {\em resource} is central to Resource Oriented Architecture (ROA). ROA is a structural design that fulfills design criteria presented by REST~\cite{rwsbook}. A resource is something that can be referred to and can have an address.  Any important information in a service interface is
exposed as a resource.

We have used the term \emph{resource definition} to define resource entity such that its instances are called resources. This is  analogous to the relationship between a \emph{class} and its \emph{objects} in object oriented paradigm. 

In our resource model, we represent \emph{resource definitions} as classes. A resource is an instance of a \emph{resource definition}, analogous to the object of a class. A collection \emph{resource definition} is represented by a class with no attributes and a normal \emph{resource definition} has one or more attributes. Each association has a name and minimum and maximum cardinalities. These cardinalities define the minimum and maximum number of resources that can be part of the association.  

\begin{figure}[h]
\centering
\includegraphics[scale=0.8]{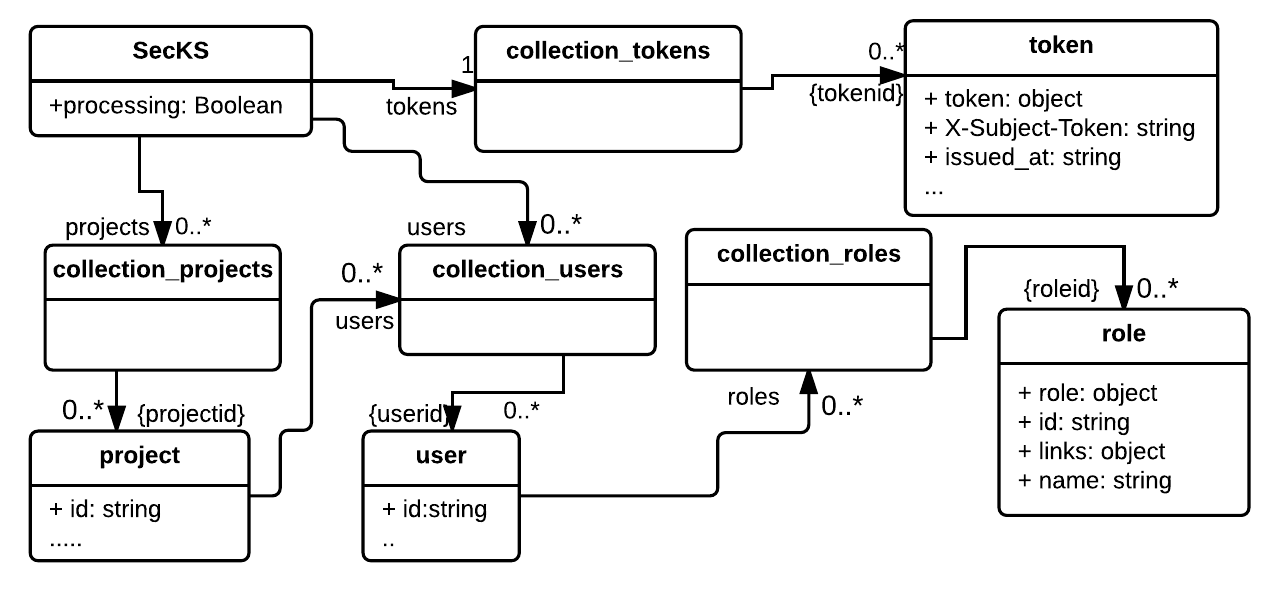}
\caption{Resource Model for KS Security Wrapper(SecKS)}
\label{fig:ks_rm}
\end{figure}

In Figure \ref{fig:ks_rm}, there are four collection \emph{resource definitions}, i.e., \textit{projects}, \textit{tokens}, \textit{users} and \textit{roles}, and five normal \emph{resource definition}, i.e., \textit{SecKS}, \textit{token}, \textit{role}, \textit{user} and \textit{project} where \textit{SecKS} represents our wrapper program capable of invoking resources in KeyStone. A collection \emph{resource definition} is represented by classes that have no attributes and  their name starts with $collection\_$.  It has one outgoing transition with multiplicity of 0...* for the contained \emph{resource definition} indicating that a collection resource can have none or many resources. A GET method on a collection resource returns a list of all the child resources it contains

We require that every association must have a role name in order to form URI addresses.The attributes of classes must be public since the representation of a resource is available for manipulation and they must have a type since they represent a document containing information of the resource, i.e. an XML document or a JSON serialized object.

\subsection{Behavioral Model: }
\label{sec:bm}

\begin{figure}[h]
\centering
\includegraphics[scale=0.8]{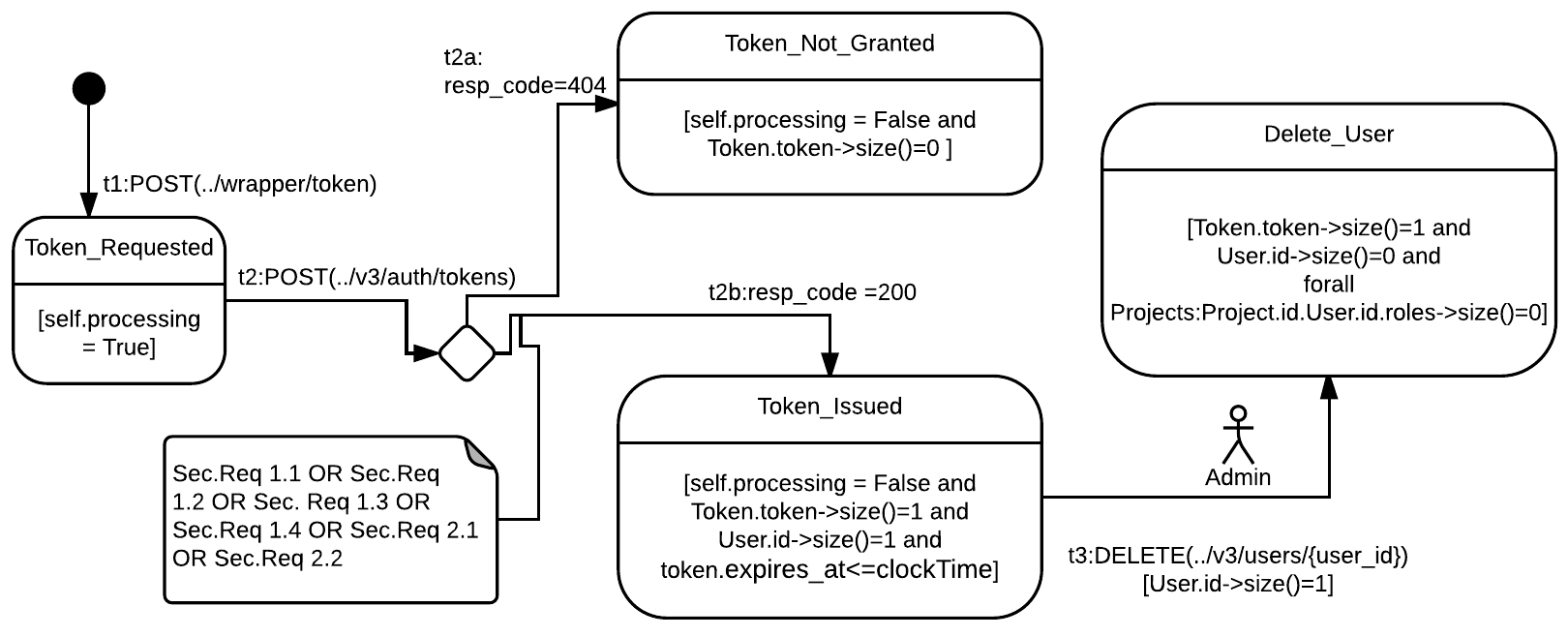}
\caption{ Behavioral Model for KS Security Wrapper(SecKS)}
\label{fig:ks_bm}
\end{figure}

The purpose of the behavioral model is to describe the dynamic structure of the behavioral interface of a REST service and is represented by a UML state-machine. Figure \ref{fig:ks_bm}  shows an excerpt of the behavioral interface of KeyStone with wrapper and provides information on what methods a user can invoke on a resource and under what circumstances.  Any client can invoke the service to request the token but only an \emph{admin} user (shown as an actor) can delete a user. Only if the client is valid, the token is generated.  

A UML state-machine has transitions that are triggered by method calls and each state has a \emph{state invariant}. State invariant is a boolean condition that evaluates to true when the service is in that particular state. Otherwise it evaluates to false. 

In our work, we define the invariant of a state using OCL \cite{OMG_OCL2} as a boolean expression over addressable resources. In this way, the stateless nature of REST remains uncompromised since no hidden information about the state of the service is being kept between method calls. 

In Figure \ref{fig:ks_bm}, state invariant for state $Token\_Not\_$ $Granted$ is written as an OCL expression: \\ $Token.token->size()=0\ \ \ and\ \ \  self.processing = False$. Here,  $Token.token->size()=0$ implies that the response for invoking GET on token resource was not 200, meaning either the resource does not exist or is not reachable to infer anything about its state. Similarly, an OCL expression $Token.token->size()=1\ $ implies the response for invoking GET on token resource was 200, meaning the resource exists. The state invariant:\\ 
$[self.processing = False\ \ \ and\ \ \ Token.token->size()=1]\ \ \ and\ \ \ User.id->size()=1\ \ \ and\ \ \ \\
token.expires\_at <= clockTime$ for $Token\_Issued$ specify that whenever a token is requested, a token is issued if the authorized user exists in the database, expiration time of token is less than the current time of the system and the wrapper is not processing the request (an asynchronous call from wrapper to KeyStone). Thus, in order to define state with stateless REST protocol, we define the state invariant as a predicate over resources.  

In addition, we constrain our behavioral model to have only side-effect methods, i.e., PUT, POST and DELETE methods as method calls for a transition. This is because only these HTTP methods are capable of making any changes to resources.

\section{Generating Contracts from SecReUM}

\label{sec:contract}
 The stateful behavior of a software requires a certain order of method invocation. These condition under which the methods can be invoked are called contracts, i.e., the pre- and post-conditions of a method. This information together with the expected effect of an operation become part of the behavioral interface of a service. Our design approach preserves the sequence of method invocations and contains behavioral information specifying the conditions under which these methods can be invoked.

\subsection{Method Contract with Functional Requirements}

The method contracts can be generated from the behavioral model. The precondition of a method should be true in order to fire the method in the behavioral model because it defines the conditions under which a method is allowed to be invoked by the client. We say that if a method \emph{m} triggers a transition \emph{t} in a state machine, then the precondition for method \emph{m} is true if the invariant of the source state of transition \emph{t} and the guard on \emph{t} is true. The post-condition constraints the implementation to provide the functionality expected from it as specified in its specification document. Thus, the post-condition states that if the precondition for invoking a method is true then its post-condition should also be true. We say, that the postcondition of method \emph{m} is true if the conjunction of state invariant of the target state of \emph{t} and the effect on transition \emph{t} is true provided its pre-condition is true. The implication principle encompasses the stateful behavior since the same method can be fired from different states of the system and have different results. Thus, if the method is fired with certain pre-conditions then the corresponding post-condition for that method should be true. 

The re-evaluation of the precondition of a method for evaluating the post-condition may not return the same values, i.e., before the method execution, since after the method execution values of some of the resources may change. This situation is kept safe by saving the resource values before method execution in local values in the wrapper. The values of these variables are later used to calculate the post-condition. We believe this is not computationally expensive as we do not need to save the copy of the whole resource/s but only the values that constitute guards and invariants that are enabled. Usually, that only requires a few bits of storage per method.

For detailed description on how contracts are generated from state-machines under different scenarios, readers are referred to \cite{porres2010nondeterministic}. 

\subsection{Security Requirements in OCL}
\label{sec:sreq1}

%Security requirements specify 
The security requirements are usually specified by security experts. We assume that they are represented in tabular format for each method. Security specifications are then translated to OCL manually. These OCL-based security requirements become part of method contract during code transformation process as shown in section \ref{sec:sreq}. 

The functional and security requirements for Keystone at the application level are not clearly separable. This is because the KeyStone functionality is to validate the identity of the user, his roles, and access rights before generating scoped or unscoped token. The security requirements on KeyStone also impose the same semantics. We classify them under security requirements since the security experts expect these behaviors from KeyStone at the application level to assure its security. We explain our approach with two important security concerns, authentication and authorization. Authentication is explained with transition \emph{t2} and authorization is explained with transition \emph{t3}. 

\subsubsection{Authentication} Authentication is an important security concern that requires that only the user with the right credentials is able to enter the system. It is also considered as one of the top three security concerns addressed by the existing model-driven security engineering approaches \cite{nguyen2015extensive}.  In Figure \ref{fig:ks_bm}, an authentication request to KeyStone triggers transition t2. The security requirements attached to t2 are listed in Table \ref{table:req}.

\begin{table}
\caption{Requirements for Authentication in KeyStone (excerpt)}
\centering
%\begin{tabular}{|>{\tiny}l|>{\tiny}p{5.5cm}| }
\begin{tabular}{|l|l| p{4cm}| }
\hline
 \textbf{No.} & \textbf{If} & \textbf{Then} \\
\hline
 1.1& User is valid and has not given  & an unscoped token should be generated\\
& scope information &  \\
 \cline{1-2}
 1.2& User is valid and has explicitly requested &  \\
  & unscoped token&  \\
  \cline{1-2}
 1.3& Token is valid and has not given & \\
& scope information &   \\ 
 \cline{1-2}
 1.4& Token is valid and has explicitly  & \\
  &  requested unscoped token &\\

  \hline
  2.1 & User is valid and has valid scope information & a scoped token should be generated   \\
  \cline{1-2}
  2.2 & Token is valid and has valid scope information &  \\
  \hline
\end{tabular}
\label{table:req}
\end{table}

These security requirements are written in OCL. For example, the security requirement for scoped token is written as:

\begin{lstlisting}
((user.credential->size()=1 or  token.token->size()=1)  and 
(request.scope->size()=1 and  not  request.scope.oclIsInvalid())) ==> (token.token->size()=1) and token.catalog->size()=1)
\end{lstlisting}

 In Table \ref{table:req}, the security requirements specify different conditions under which scoped and unscoped tokens are issued and are written in the if-else format on resources and resource attributes. The security requirements can also be in a statement form enforcing some rule, for example, the authorization requirement explained in the next section.

\subsubsection{Authorization}
 
Authorization defines access rights of users by defining permissions on the user, user roles, and user groups. KeyStone determines whether a request from the user should be accepted based on the policy rules defined in Role Based Access Control (RBAC). In Figure \ref{fig:ks_bm}, \emph{t3} can only be fired by an \emph{admin} user. In addition, the guard value shows that the user to be deleted should have initially existed in the system. The information about actors in the behavioral model can be realized in two ways. 
\\
% \\
% \begin{lstlisting}
% ((user.id->size()=1 \quad and \quad user.role='admin'))
% \end{lstlisting}
% \\
1) The developer can use this information to implement the access rights on resources and help users in understanding and writing correct authorization headers. Different authentication mechanisms can be implemented to control access to resources \cite{authLink}. If the Basic authentication mechanism is implemented, the client sends the username and password to the server in the authorization header. The authentication information is in base-64 encoding. It should only be used with HTTPS, as the password can be easily captured and reused over HTTP.

%  In a typical setting, the authorization header is constructed by first combining \emph{username} and \emph{password} into a string "username:password" and then encoded in based64. A typical authorization header in Basic authentication is shown below:

%  {\small
% \begin{lstlisting}
% DELETE /v3/users/22/ HTTP/1.1
% Host:http://localhost:5000/v3/
% Authorization: Basic aHR0cHdhdGNoOmY=
% \end{lstlisting}
% }

% In case an anonymous requests for a protected resource, HTTP can enforce basic authentication by rejecting the request with a 401 (Access Denied) status code.

% {\small
% \begin{lstlisting}
% HTTP/1.1 401 Access Denied
% WWW-Authenticate: Basic realm="User"
% Content-Length: 0
% \end{lstlisting}
% }

For KeyStone, authorization to resources is checked with \emph{token}. A typical call from \emph{curl} to access \emph{User} resource using user's \emph{token} is given as:

\begin{lstlisting}
curl -s \ -H "X-Auth-Token: $OS_TOKEN" \ "http://localhost:5000/v3/users" 
 \end{lstlisting}

% 2) The security requirements can be attached as predicates of boolean variables to transitions and translated to code as such. All the boolean variables for security requirements are initialized to be false, e.g. $sreq1=False$. Whenever, the postcondition of a requirement is true in the implementation, the boolean variable is set as True, $sreq1=True$. The boolean values of these security requirements are displayed to the user after the system is tested with different test cases. This added feature gives  clear information to security experts as to what security requirements are satisfied and in identifying the met and unmet security requirements by the system without looking into the implementation details.

2) It becomes part of the method contract. The security requirement for the authorization is: \emph{Only an admin user can delete a user.} In OCL, it is written as: $user.role='admin'$. 

This can be specified in UML as notes (not shown in Figure \ref{fig:ks_bm}
 due to space limitation). In the next section, we define rules on how they become part of the method contract.

\subsection{Method Contracts with Functional and Security Requirements}
\label{sec:sreq}

The security requirements are merged with functional requirements during the translation process to code. 
In our example, the KeyStone service is invoked by POST method on the token resource \\
($POST(../v3/auth/tokens)$). We populate our definition of contracts with security requirements given above such that:
\begin{itemize}
    \item The statement in \emph{if} clause becomes part of the method pre-condition
    \item The statement in \emph{else} clause become part of the method post-condition
    \item The statement/s that are not part of \emph{if-else} clause become part of both the pre- and post-conditions. By checking the rule in pre-condition, the user request is validated before processing the method and causing undesired changed in the system. By placing in the post-condition, the system is validated that it behaves as expected and without side effects. This serves as a double check on security requirements.
\end{itemize}
 We, thus, require that for KeyStone to generate a token, the following method contract must be met:
 \\
%\tdi{Need to updated post-condition in listing}

{
\begin{lstlisting}
PreCondition(POST(../v3/auth/tokens)):

[(self.processing = True and (user.credential->size()=1 or
token.token->size()=1 and token.expires_at-> <= clockTime]) 
and 
((request.scope->size()=1 and request.scope <> 'unscope' and not request.scope.oclIsInvalid())
or (request.scope->size()=0 or request.scope.oclIsInvalid() or 
request.scope = 'unscope' ))]

PostCondition(POST(../v3/auth/tokens)):
[((user.credential->size()=1 or
token.token->size()=1 and User.id->size()=1 and 
token.expires\_at <= clockTime$) and 
((request.scope->size()=1 and request.scope <> 'unscope' and not request.scope.oclIsInvalid())==>
(self.processing = False and token.token->size()=1 and token.catalog->size()=1))
or ((self.processing = True and request.scope->size()=0 or request.scope.oclIsInvalid() or 
request.scope = 'unscope') ==> (self.processing = False and token.token->size()=1 and token.catalog->size()=0))]

\end{lstlisting}
}

The preconditions in the listing above show the boolean expression that should be true for invoking a POST on KeyStone for either scoped or unscoped token. The postcondition circumscribes different scenarios for scoped and unscoped token. In order to return an unscoped/ scoped token, the previous values, i.e. the values before method invocation, are checked. If the previous values require an unscoped/ scoped token then the response of method calls are checked to ensure if unscoped/ scoped token is actually delivered. The previous values, i.e., the values before the method invocation are stored as local variables in the wrapper program. 

For authorization, the method contract for DELETE on user resources is given as:

{
\begin{lstlisting}
PreCondition(DELETE(../v3/users/{user_id}))):

[self.processing = False and token.token->size()=1 and 
user.id->size()=1 and token.expires_at <= clockTime 
and user.role='admin']

PostCondition(DELETE(../v3/users/{user_id})):
[(self.processing = False and token.token->size()=1 and 
user.id->size()=1 and token.expires_at <= clockTime 
and user.role='admin') ==>
(token.token->size()=1 and user.role='admin' and 
user.id->size()=0)]
\end{lstlisting}
}

In this listing, $user.role='admin'$ is checked before invoking DELETE method on $User$ resource to ensure that user with the right credentials is making the desired change in the system. Interestingly, $user.role='admin'$ is also a part of the post-condition, i.e., the credentials of the user are checked before and after the method execution to ensure that the system change is made by the right user. This double check of the security requirement for authorization provides added security and guards the system against the malicious user during the communication.

\section{Implementation of OpenStack and a Service Monitor}
\label{sec:imp}
 
We deployed OpenStack on a separate machine as single node deployment using DevStack. The machine had UBuntu 16.04 installed with 8GB RAM and i3 processor. The Keystone service was invoked using OpenStack client and \textit{curl} commands over the network using a machine with MacOS and 8 GB RAM.  We implemented our monitoring mechanism in Django \cite{holovaty2009definitive} by using the behavioral and security information present in our design model. At a glance, Django can be understood with its three basic files that support separation of concerns, i.e. models.py, urls.py and views.py where models.py contain descriptions of database tables, views.py contains the business logic and urls.py specify which URIs map to which view. For a detailed working of Django Framework, readers are encouraged to read Django Documentation~\cite{djdoc} and Django Book~\cite{djbook}.

The service monitor is implemented as a service proxy (wrapper). It listens for requests from the cloud user, verifies the conditions to invoke the method and then forward it to the actual service implementation. 

A service monitor can be used to continuously verify the functionality of an implemented cloud service. This monitoring mechanism checks that the open source cloud environment continues to follow its security concerns despite frequent updates in the code by other developers. We consider our implementation of monitoring mechanism as a complementary approach to other security validation mechanisms. 

In the current implementation, we validate security concerns of authentication and authorization. If the cloud user does not invoke the cloud service with right credentials, the error messages are returned back without invoking the cloud service. If the pre-conditions for invoking the method are met, the method is invoked on the cloud service and the response is received from the cloud. The implemented wrapper then goes through the response and verifies if the response from the cloud service is as expected, i.e. according to the user's request. For example, if the user has given credentials for the unscoped token, then the cloud service should provide the unscoped token correspondingly. If the scoped token is requested, then the cloud service should return the scoped token. Thus, the cloud user is checked for an invocation to the service under right conditions and the cloud service is constraint to provide the implementation as specified.

The main steps in our implementation phase are:
\begin{itemize}
  \item Implement database tables in models.py
  \item Create views for each resource and its transitions in views.py
  \item Map relative URIs from resource model to respective views in urls.py.
\end{itemize}

Our models.py contain only one class \textit{kswrapper}, i.e., the wrapper class as shown in the listing below since our wrapper saves all the information required for processing requests. The other required pieces of information are retrieved from the open source implementations through their REST APIs at runtime.

\begin{lstlisting}[language=Python, caption=Implementation of Database Model for KS wrapper, label=lst_models, language=Python]
from django.db import models

class kswrapper(models.Model):
    ksDate =  models.DateTimeField()
    tokenId = models.CharField(max_length=200)
\end{lstlisting}

In the second step, a view is defined for each resource in our resource model. These views contain information on allowed and not-allowed methods on resources retrieved from behavioral model and also the contracts for these invocations. The incoming request to the view is verified against the allowed methods and redirected to the view that supports the request method for the resource.

In a proxy interface for the KeyStone service, a POST method on the token resource for a scoped token is implemented as:

\begin{lstlisting}[language=Python,caption= Excerpt of POST view in Proxy Interface for POST on \textit{Token} resource, label=lst_proxyget]
def ks_token(request, body):
   	if not request.method in ["GET", "POST"]:
        return HttpResponseNotAllowed(["GET", "POST"])
    if request.method == "GET":
        body1 = body
        return ks_token_get(request, body1)
    if request.method == "POST":
        body1 = body
        return ks_token_post(request, body1)

def ks_token_post(request, body):
   	parsed_json = json.loads(body)
   	sc_var=False
   	scopeVar= None
   	try:
   	   	scopeVar=parsed_json["auth"]["scope"]
   	   	if scopeVar:
   	   	   	sc_var=True
   	except KeyError:
   	   	print "The object does not have scope information"
   	   	sc_var=False
   	   	scopeVar= None
   	req_method=parsed_json["auth"]["identity"]["methods"][0]
   	if req_method == "password":
   	   	uname=parsed_json["auth"]["identity"]["password"]["user"]["name"]
   	   	un_flag=True # This means that request has username information
   	else:
   	   	if req_method=="token":	    		
   	   	   	tid = parsed_json["auth"]["identity"]["methods"][0]
   	   	   	token_flag=True # This means that request has token information
   	req = urllib2.Request('http://130.232.85.9/identity/v3/auth/tokens',body)
   	processing=True
   	req.add_header("Content-Type",'application/json')
   	if (processing==True and (un_flag == True or token_flag==True) and ((sc_var== True  and scopeVar != "unscope") or (scopeVar== False or scopeVar != "unscope"))):
   	   	response = urllib2.urlopen(req)
   	   	the_page = response.read()
   	   	processing=False
   	   	parsed_json2 = json.loads(the_page)
   	   	cat_var= False
   	   	try:
   	   	   	cat=parsed_json2["token"]["catalog"]
		   	if cat:
   	   	   	   	cat_var=True
   	   	except KeyError:
   	   	   	print "The object doesn't have catalog information"
   	   	   	cat_var= False
   	   	response_headers = response.info()
   	   ##skipping the extraction of other attributes like expires_at for token
   	   	token = response.info().getheader('X-Subject-Token') # getting token 	
   	   	body= "token:", token
   	if ((response.code == 200 or response.code == 201 and p==False  and notExpired=True) and ((un_flag == True or token_flag==True) and (sc_var== True  and scopeVar != "unscope")) and (cat_var==True )):
   	   	# for scoped token
   	   	T_type="T_type = Scoped Token: "
   	   	body= str(body) + T_type
   	   	r = HttpResponse(the_page)
   	   	response = HttpResponse(body)
   	   	response.status_code = 200
   	   	return response
   	elif ((response.code == 200 or response.code == 201 and p==False and notExpired=True) and ((un_flag == True or token_flag==True) and (sc_var== False or scopeVar == "unscope")) and (cat_var==False )):
   	   	# for UNscoped Token
   	   	T_type="T_type = UNscoped Token: "
   	   	body= str(body) + T_type
   	   	r = HttpResponse(the_page)
   	   	response = HttpResponse(body)
   	   	response.status_code = 200
   	   	return response
   	else:
   	   	print response.code
   	   	response = HttpResponse()
   	   	response.status_code = 404
   	   	return response			
\end{lstlisting}

The listing above follows the following algorithm: 

1) When an HTTP request comes for authentication to \textit{token} view (associated with token resource), the request is filtered according to request method and redirected to the corresponding \textit{view}. In the listing above, we only show \textit{post} view in detail.

2) The request body is parsed and checked to see if it has scope values. The scope flag (\textit{sc\_var}) is set to \textit{true} if scope values are present and otherwise \textit{false}. 

3) If the preconditions are satisfied, then the Keystone REST API is invoked with the authorization header.

4) The response from the Keystone is parsed to check if it contains right information, e.g., does it have catalog information. The response body from KeyStone for POST request on \textit{token} contain catalog information if it is a scoped token.

5) Different variable values, some of which were set earlier by parsing the request body before method invocation to KeyStone API and some that were set by parsing response body after method invocation to KeyStone API, are combined together in boolean expressions as explained earlier in section \ref{sec:sreq}. Based on these conditions, successful or unsuccessful responses are given by the wrapper program.

In the real proxy interface, i.e. our wrapper, a method is implemented for each of the selected methods that are invoked on the Keystone component of OpenStack using urllib2. urllib2 is a python module that is used to fetch URLs~\cite{urllib}.  The number of methods that are selected for implementation in the wrapper can be reduced by selecting only those resources that are considered assets by security experts or any other priority criteria. We leave the job of selecting the resources and methods to be implemented by the wrapper program on security experts and quality experts based on their priority lists.

In the third step, the relative URIs shown in the resource model are mapped to the respective views. Every resource in our resource model is addressable. We can get the relative URI for each resource directly from Figure \ref{fig:ks_rm} that is then mapped to the respective views as shown in Listing \ref{lst_uris}.

\begin{lstlisting}[language=Python,caption=Relative URIs and views mapping for KSWrapper, label=lst_uris]

urlpatterns = [
   url(r'^kswrapper/', views.index, name='index'),
    url(r'^kswrapper/tokens/',  views.ks_tokens, name='ks_token'),
    url(r'^domains/', views.domains, name='domains_get'),
    url(r'^admin/', admin.site.urls),
]

\end{lstlisting}

%Each GET view returns an HTTP response object. When a POST, PUT or DELETE method is implemented in the proxy interface, it manipulates the status codes of the HTTP response objects and asserts them as method pre and post conditions. The other methods are not shown here due to space limitation.
\section{Applications of the Approach}
Cloud security is generally associated with the use of latest technologies and security techniques to protect applications, data and infrastructure associated with cloud computing. However, as security technologies and techniques evolve so do the techniques used by the attackers to attack the clouds. There is a need to consider security of cloud at the software and application level in addition to the use of latest security-related technologies. 

By using models to define behavioral interfaces for REST APIs and the approach described in this article to generate contracts as code skeletons in a wrapper program, we can benefit from previous and future efforts in test case generation from behavioral contracts while using a familiar and standardized visual notation. In addition, our work complements the security technologies in providing secure cloud by providing a continuous validation and monitoring approach for the security concerns. 

The security concerns of authentication and authorization are part of behavioral contracts implemented in the behavioral wrapper. The Identity service, like KeyStone of OpenStack, define role assignments, i.e. what role does a user has on a specific project or domain \cite{pepple2011deploying}. However, the capabilities of roles, i.e. what can or cannot these roles do is defined in the authorization policies defined separately for each service in OpenStack. Different authorization policies can be defined for different services in the same cloud environment. Using our approach, the specification of different authorization policies along with the functional contracts can become part of behavior wrapper which can be used to validate the security concerns in the actual cloud implementation. The contracts can thus be exploited for the generation of test oracles and test cases can validate the implementation of a cloud. Test oracles are used to determine whether a test has passed or failed. In the context of test case generation and test oracle generation, we can take advantage of several efforts done previously to validate the behavior of classes and services using contracts \cite{ciupa2005automatic} \cite{dai2007contract}.

The wrapper program with behavioral specifications can also be added as a proxy interface to the implemented cloud to monitor its functioning and security compliance. This facilitates location of the fault in and application by observing the conditions that are not being met and by which methods. It can also check for any failure caused by a network fault, late delivery or if an implementation violates a certain pre or post condition of a method.

Finally, cloud designers and developers along with security experts can use the models with usage scenarios and security information as detailed documentation on how to use a cloud correctly.

\section{Related Work}

Research in using models to develop and analyze secure systems has been an active area of research for more than a decade.The work of Nguyen et al. \cite{nguyen2015extensive} provides a comprehensive review of efforts done in the area of model-driven development of secure systems. Their work encompasses various modeling approaches like UML-based approaches, UML profiles, DSLs and aspect-oriented approaches and analyzes them for their support for model-to-code and model-to-model transformations, verification, validation and different types of security concerns. UML has been used much to model security concerns. Some approaches use only UML (e.g., \cite{abramov2012methodology}, \emph{MDSE@R} \cite{almorsy2014adaptable}, \emph{AOMSec} \cite{georg2009aspect} etc.) and some use UML profiles(e.g., SECTET \cite{alam2004model},\emph{UMLsec}\cite{jurjens2001towards}, etc.)

 In \cite{abramov2012methodology}, Abramov et. al. present a model-driven approach to integrate access control policies on database development. SECTET \cite{alam2004model} provides a model-driven security approach for web services. They also use OCL to define constraints on UML to provide access control. The approach generates XACML policy files that provide a platform-independent policy for enforcing the access control policy. The SECTET framework mainly addresses authorization and provides state-dependent permissions that are not applicable to REST interfaces.  \emph{UMLsec}\cite{jurjens2001towards,jurjens2007tools} provides a comprehensive and consistently progressing approach to formally analyze the security properties. \emph{MDSE@R} \cite{almorsy2014adaptable} provides a UML profile based approach that uses aspect-oriented programming to integrate security concerns at the runtime. \emph{AOMSec} \cite{georg2009aspect} also uses aspect-oriented approach to model security mechanism and attacks to the system. A detailed analysis of existing literature is out of the scope of this paper. However, compared to previous work our work strongly relies on existing UML without the need of any new profiles. This gives the benefit of using many well-known and mature tools with a wide user base for our approach. Our work also caters well with the stateless nature of REST. 

In our previous work \cite{troubitsyna2016towards},\cite{troubitsyna2016integrated}, we have investigated the problem of deriving the security requirements from the formal system model. The approach presented in this paper, complements this work by bridging the gap  between the security requirements and the actual code.  In \cite{laibinis2016formal}, we have proposed a method for identifying security vulnerabilities using formal architectural model. The UML modeling patterns proposed in this paper can significantly facilitate constructing such models and hence, enable the industrial adoption of the proposed technique.

%\cite{cui2015security} - keystone security analysis

%Researchers have used different types of models to model security concerns. Primarily, we found following modeling strategies used widely, \emph{aspect-oriented modeling}, \emph{UML diagrams}, UML profiles or DSLs. 
%\tdi{Start from skimming on: Model-driven security testing from \cite{nguyen2015extensive}}
\section{Conclusions}
 
Open source cloud frameworks are becoming popular as more and more enterprises are opting for private clouds for their work amid data and network security concerns.  Security experts are often looking out for ways to assure that their security expectations from a system are met.  Our approach provides security experts with a model-driven approach that facilitates them by providing a semi-automatable approach for validating the open source cloud environment for its security concerns like authentication and authorization. We show how the security concerns can be integrated into the behavioral models of REST services and how method contracts can be generated from them that can be later used to validate any security loopholes in the open source software in case of frequent updates. The approach is applied on the KeyStone component of OpenStack.

We have presented in detail the implementation of service monitor using Django web framework. We also present the applications of our behavioral and security modeling approach along with the contract generation methodology for cloud security. In our future work, we plan to further extend our work for different authorization scenarios and validating cloud implementations for their security concerns.

\bibliographystyle{eptcs}
\bibliography{fmMdd17}
%\bibstyle{plain}

\end{document}